# Brain Energetics, Mitochondria, and Traumatic Brain Injury

Haym Benaroya, Rutgers University
benaroya@rutgers.edu

Abstract

We review current thinking about, and draw connections between, brain energetics and metabolism, mitochondria and traumatic brain injury. In addition to summarizing current thinking in these disciplines, our goal is to suggest a framework for mechanisms and pathways based on optimal energetic decisions.

# General Introduction

As neuroscience makes rapid progress in mapping causes and effects at subcellular, cellular, tissue and system levels, the time may be ripe for increased efforts at the mathematical modeling of these causes and effects. Descriptive understanding is extremely valuable, but these tend to be snapshots of coupled and time-dependent processes simultaneously governed by dozens of parameters. Mathematical models, where possible, offer dynamical equations that govern these coupled time-dependent (and perhaps space-dependent) parameters, some if not all of which are stochastic in nature, and apparently sensitive to parameter variations. While these possible mathematical equations are data-dependent, they are not data-driven. They permit computational analysis, parameter variations, and sensitivity analyses. They can be used to estimate the outcomes of clinical interventions, and for experimental design, perhaps reducing the number of these in time. This review suggests opportunities for such mathematical models where optimal decisions are apparently made in certain energetic pathways.

    Whereas neurons and glia are the components of brain function, energy homeostasis must be maintained to assure proper functioning. This homeostasis is the product of metabolic reactions that are coupled to energy demands in space and time throughout the brain, and regulated by feedforward and feedback mechanisms. The mitochondria are at the center of this universe, providing up to about 95% of this

energy, where and when it is in demand in the central nervous system. Any mismatch between supply and demand over a significant time interval invariably initiates cascades of dysfunction leading to well-known neurodegenerative and neuropsychiatric pathologies. Energy is the currency of brain function, and is likely the basis for the physiological or pathological pathway decisions taken.

This paper summarizes our current understanding of the coupling between brain energetics (and metabolism), the mitochondria, and traumatic brain injury. Our summary is unavoidably selective given the vast literature in all these fields. We have tried to capture an essence of what present-day primarily experimental evidence implies regarding these functions. Some hypotheses are suggested intermittently and at the end of the paper.

# Energetics

1. Introduction

Among placental animals, human brain size relative to body size is the largest [Gibbons 1988]. Brain activity requires significant energy. To understand brain functioning, cellular and molecular mechanisms of brain activity need to be identified and coupled to energy metabolism. This approach has been understood for over a century [Magistretti et al 1999] due to the tight coupling between neuronal activity and energy metabolism. Imaging technologies have permitted this mapping. Magistretti and Pellerin [1999] propose what came to be known as the astrocyte-neuron lactate shuttle hypothesis (ANLS), where lactate may be the preferred energy substrate of activated neurons under aerobic conditions. To this day there are debates on this hypothesis, and these will be introduced as we proceed. Key components of the above-mentioned coupling are the neurotransmitter glutamate and astrocytes, a glial cell. Chih et al [2001] pick up the discussion on the so-called astrocyte-neuron lactate shuttle, where glial energy needs are met by anaerobic glycolysis, while neuronal metabolism is fueled by lactate released from the glia during neuronal activity. Based on enzyme kinetics and substrate availability, it is suggested that neurons likely use ambient glucose as the major substrate during activity rather than glial-derived lactate. Furthermore, it is possible that lactate might be used after neural activity when glycolytic rates slow, and during recovery from pathological insults. This possibility may be tied to glial behavioral changes after traumatic brain injury. Magistretti and Allaman [2016] provide an updated overview.

As discussed in detail subsequently, a key component of the neuronal energy equation is the organelle, the mitochondrion. As we find out, mitochondria transport to locations within the cell where energy is required. For the neuron, this means that it must ensure an adequate supply and distribution of mitochondria that are healthy and functional throughout its axon and dendritic arborizations. Given the morphological complexity of these cells, and that neuronal function is closely coupled with glial cells, it is possible that evolutionarily conserved optimization mechanisms have significant roles in assuring sufficient energy as needed.

2. Astrocytes

Astrocytes possess specialized processes that cover the surface of intraparenchymal capillaries, where they can control glucose uptake, as well as processes that wrap around neuronal synapses, sites of receptors and reuptake of neurotransmitters. Such a strategic presence allows astrocytes to sense synaptic activity and couple these with energy metabolism. A review of energy metabolism in astrocytes by Hertz et al [2007] underscores their importance to brain function. Primary roles are astrocytic interactions with the vasculature, neurons, and other astrocytes via signaling, the regulation of blood flow, the modulation of impulse transmission, and the synthesis and degradation of glucose-derived neurotransmitters, which modulate local energy metabolism in the brain. All these roles are dependent on a connection to various metabolic pathways, thus coupling astrocytic functions to energetics and metabolic fluxes. It is noted that under rest conditions, astrocytes account for approximately 30% of oxidative metabolism in the brain and about 50% of glycolysis, but very little glycogenolysis. This suggests that there are functional activities that are not known or fully understood. Any disruption of these processes rapidly alters neuronal function, suggesting mechanisms for gliosis after brain insult.

The typical astrocyte cell volume is mostly distributed in the processes, with about 2% of the volume in the cell body and about 60% for the organelle-containing processes. Oxidative capabilities rest within the cell body and larger cell processes that contain the mitochondria. The remaining almost 40% of astrocytic volume is contained in the threadlike filopodia, and sheet-like lamellipodia, which are highly mobile but cannot generally accommodate mitochondria, which are about twice as wide (although there is evidence that mitochondria exist in the fine astrocytic processes). However, these small extensions respond rapidly to increases in energy demand via glycolytically-derived ATP [Hertz et al 2007].

Understanding how neurons and astrocytes regulate blood flow is evolving, as this knowledge is critical to the ability to develop new therapeutic approaches. It appears that neurotransmitter-mediated signaling, in particular by glutamate, rather than energy use, has a key role in regulating cerebral blood flow, where astrocytes play a key role. It is significant that astrocytes take up over 90% of the glutamate released as a neurotransmitter. If this did not occur, and the glutamate was taken away by the blood, then the glucose supply to the brain would have to be approximately doubled in order to supply the carbon skeletons needed for glutamate synthesis, as well as the glucose needed as a substrate for ATP production [Nortley and Attwell 2017]. Signaling pathway differences can lead to different blood flow control in various brain areas, and this control occurs at the capillary and arteriole levels, where $O_2$ concentrations regulate the relative importance of the signaling pathways involved [Attwell et al 2010].

Peppiatt and Attwell [2004] report on the importance of astrocytes to blood flow regulation. These glial cells simultaneously surround neurons with their cell membrane and have extensions, called endfeet, close to capillaries. A rise in calcium levels may either dilate or constrict arterioles and increase or reduce local blood flow, depending on other factors at play.

There are approximately as many astrocytes as there are neurons in the brain. Astrocyte processes cover approximately 63% of capillaries, with the remaining area almost covered by pericytes, which are spatially isolated contractile cells that are believed to be responsible for a major fraction of increased blood flow [Nortley and Attwell 2017]. It has been implied that low levels of neuronal activity lead to an increase in blood flow solely by dilating arterioles (possibly mediated by NO without intercession by either astrocytes or pericytes), while higher activity also dilates capillaries via astrocytes and pericytes. That glycolysis in not purely astrocytic, and that lactate-fuelled oxidative phosphorylation is not purely neuronal, removes some of the basis for the ANLS hypothesis, but there are no firm conclusions to date [Nortley and Attwell 2017].

Escartin et al [2006] review the contributions of NMR to astrocyte-neuron coupling in the regulation of brain energy metabolism. The dialogue between astrocytes and neurons is a critical aspect of proper brain function. The brain has significant and continuous energy needs and, especially in periods of high activity and therefore demand, it is highly vulnerable to a drop in supply, that is, energetic dysfunction, and dysfunction of homeostasis. Such dysfunction, as well as improper recycling and metabolism of glutamate, are at the core of numerous chronic neurodegenerative diseases, in particular the big-four diseases (APHALS): Alzheimer's, Parkinson's, and Huntington's, and amyotrophic lateral sclerosis (ALS).

Astrocytes have been found to be a central player in brain function on many levels. In addition to supporting neuronal function through the supply of energy metabolites and by recycling neurotransmitters, they play a central role in brain homeostasis (including during sleep). That is, glutamate, ion and water homeostasis, oxidative stress defense, energy storage as glycogen (that can be rapidly metabolized without ATP), scar formation and tissue repair (as a response to insult and traumatic brain injury), release of gliotransmitters for synaptic activity modulation, and synaptic formation and remodeling (plasticity). Additionally, they have roles in memory consolidation and the regulation of breathing. Astrocytes are a morphologically heterogeneous population, with a variety of receptors, ion channels and proteins. In some instances their effects are negative, as in reactive astrogliosis and the formation of a glial scar, where the fine line between neuroprotection and neurotoxicity is not well understood. Given the importance of astrocytes on metabolic function, and that metabolic dysfunction has been shown to lead to neurodegenerative disorders, a deeper understanding of astrocytes and their couplings is critical for the development and implementation of clinical interventions [Allaman et al 2011]. One metabolic susceptibility is the appearance of astrocytoma cells, which are the most common types of cancer in the brain, and the most common gliomas [Turner and Adamson 2011].

Astrocytic gap-junctions appear to be crucial to the central role of astrocytes to neurometabolic coupling [Escartin and Rouach 2013]. The power of such coupling is not at the individual neuro-glia-vascular unit, but rather based on the networks of astrocytes connected by gap junctions. Thus, models of astroglial metabolic support of neuronal activity must encompass the gap-junction-mediated astroglial networks. (Similarly, networks of mitochondria can work as a group rather than as individual mitochondria.) Such a modeling perspective more accurately reflects their influence over local synapses, as well as distal neuronal circuits. This is useful since optimal metabolic pathways are chosen for proper energetic functioning as well as the minimization of energy expenditures. This approach is strongly suggested by the likelihood [Escartin and Rouach 2013] that metabolic waves pass through astroglial networks in both physiological and pathological conditions, enforcing metabolic coupling over extensive distances. The significance of astrocytic gap junctions has been understood for some time [Scemes and Spray 2004]. Such gap junctions between astrocytes provide pathways for direct intercellular exchanges of ions, nutrients and signaling molecules. Additionally, such pathways may be a part of the machinery for delivering nutrients to neurons, removing excess $K^+$ and glutamate from intercellular space, spreading intercellular $Ca^{++}$, and exchanging cell death signals. Gap junction connected astrocytes may be viewed as a hybrid between cellularized individual cells and syncytial cells, that is, a functional syncytium. They are individual cells, but are connected via the gap junctions that coordinately and cooperatively react to long-range neuronal activity and environmental stimuli.

Such coordination and cooperation requires biochemical energetic "choices" that are optimum in some sense, either minimizing energy cost, or maximizing speed, or some combination of these. Furthermore, it is possible that reflexive (autaptic) gap junctions in astrocytes can alter gap junction coupling strength so as to isolate or integrate microdomains in order to fine tune molecular transport throughout the astrocytic network in response to neuronal activity. This appears to correlate with the view [Barros 2010 discussed below] that to maintain steady brain energy use, regions of active cells may be surrounded by passive cells, thus localizing energy use as needed. Gap junction permeability and selectivity modulation play roles in these dynamic mechanisms. Such dynamics can only be efficiently captured by mathematical models that are representative of time- and space-dependence of mechanisms.

In pathological conditions, gap-junction-mediated networks of astrocytes can contribute to the transfer of survival signals, such as metabolic substrates, in situations where local acute pathological conditions occur in which distal neurons are no longer connected to the bloodstream. Examples include ischemia, traumatic brain injury, or infections [Escartin and Rouach 2013], in cases where the gap-junction channels are still functional.

As one of the foremost proponent of the importance of glial cells, Barres' [2008] makes a critical point in his poignant paper: "Quite possibly saving astrocytes from dying in neurological disease would be a far more effective strategy than trying to save neurons (glia already know how to save neurons, whereas neuroscientists still have no clue)." Neuroenergetics is now a study of integrated astrocyte-neuron systems where cooperation and complementary functions are critical for brain health [Bélanger et al 2011, Bouzier-Sore and Pellerin 2013]. Astrocytes have a role in higher cognitive functions, and therefore contribute to the pathogenesis of several brain disorders [Dossi et al 2018]. Neuron-astrocyte network computational models are valuable *in silico* representations [Oschmann et al 2018].

3.  Mathematical Modeling

The mathematical modeling of energetics and metabolism started in earnest with the availability of imaging data. One such effort by Aubert et al [2001] models brain functional imaging using a system of differential equations that govern the following processes that are involved in brain activation: (1) sodium membrane transport, (2) Na/K ATPase, (3) neuronal energy metabolism (i.e., glycolysis, buffering effect of phosphocreatine, and mitochondrial respiration), (4) blood-brain barrier exchanges, and (5) brain hemodynamics. The purpose of these mathematical models is to enable the interpretation of MRS and fMRI data. Brain activation and metabolism are assumed to be coupled due to either changes in ATP and ADP concentrations following activation of Na/K ATPase resulting from changes in ion concentrations, or the involvement of a second messenger such as calcium. Physiological interpretations are challenging. The above-mentioned differential equations are typically first-order rate equations. Aubert and Costalat [2002] additionally model the coupling between brain electrical activity, metabolism, and hemodynamics. These models help us with a better interpretation of functional brain imaging signals, as well as possibly relating these activities to brain diseases. These approaches are explored by Aubert et al [2002] for gliomas, which can display significant changes in concentrations of energy metabolism molecules.

Aubert and Costalat [2005] continued the exploration of the ANLS framework using a mathematical model of the energy metabolism that couples neurons and astrocytes. Their model suggests possible physiological mechanisms where ANLS is compatible under certain conditions with the classical view of brain metabolism. Aubert et al [2005] continue the mathematical modeling of brain lactate kinetics, that intraparenchymally formed lactate can be a significant energy substrate for neurons. Their model predicts that neurons are lactate-consuming cells while astrocytes are lactate-producing cells; in agreement with *in vivo* observations, such production and activation increase in a coupled way. In a continuation of their mathematical modeling of metabolic pathways and the study of the role of lactate in the relationship between astrocyte and neuron, Aubert and Costalat [2007] examine compartmentalization of brain energy metabolism, with Aubert et al [2007] providing a functional neuroimaging perspective. They discuss possible modifications of the functional relationships between glia and neurons upon activation, and under what conditions the ANLS hypothesis can work, noting that in order to understand brain energy metabolism a thorough model of the interactions between neurons and glia is fundamental.

Vatov et al [2006] construct and test a mathematical model of brain energy metabolism that incorporates the following parameters: cerebral blood flow, partial oxygen pressure, mitochondrial NADH redox state, and extracellular potassium. It aims to reproduce pathological conditions, such as complete and partial ischemia, and cortical spreading depression. This is a point-model, where only variability in time is calculated, not in space, with the potential to describe pathogenesis of tissues in various experimental and clinical situations.

4.  Energy Expenditures

The brain depends on glucose as its primary fuel due to the selective permeability of the blood-brain barrier. Glucose cannot be replaced as an energy source, but it can be supplemented during high levels of physical activity or during starvation. Neuroactive compounds such as glutamate, aspartate, and glycine, as well as other compounds required by the brain to function that cannot pass through the blood-brain barrier, need to be synthesized from glucose within the brain. Approximately 5.6 mg of glucose is consumed per 100 g of human brain tissue per minute. While action potentials are evolutionarily efficient, much of the energy used in the brain is used on synaptic activity; the human cortex requires approximately $3 \times 10^{23}$ ATP/s/m$^3$, and the energy expenditure to release one synaptic vesicle is calculated to be approximately $1.64 \times 10^5$ molecules ATP [Mergenthaler et al 2013].

    Attwell and Laughlin [2001] present detailed calculations of energy expenditures on different cellular mechanisms in the brain. While the human brain is 2% of the body's weight, it demands 20% of the body's resting metabolic output, that is, glucose-derived energy. Action potentials and postsynaptic effects of the neurotransmitter glutamate are estimated to consume much of that energy, 47% and 34%, respectively. The resting potential requires 13%, and glutamate recycling requires 3%. It is clear that the action potential rate has a significant effect on energy usage, and is, in this paper, tied to oxygen consumption rates. Various other energy consumption rates are provided in considerable detail.

    Laughlin and Attwell [2004] update their earlier energy budget. Postsynaptic energy consumption dominates the cost of synaptic transmission, roughly 85%. That mitochondria are concentrated around synapses, in dendrites and presynaptic terminals, is consistent with the view that synapses require much more energy than action potentials. The variety of homeostatic mechanisms required to reset ionic gradients after neural signaling require energy produced by the hydrolysis of ATP to ADP, where most of the brain's ATP is regenerated from ADP by the aerobic glycolysis of glucose. Interestingly, it is noted that synaptic costs can only be probabilistically characterized for the following reasons. There is a large variance in the number of

postsynaptic ion channels activated by the release of a vesicle of neurotransmitter, and this can vary between synapses. Secondly, the proportion of synapses that release neurotransmitter in response to an action potential is a function of signaling history, and also has variance. This is discussed further subsequently.

In the same paper, Laughlin and Attwell provide interesting perspectives on the efficiencies and optimalities in the evolved brain. This involves reduced metabolic costs of signal transmission and synaptic transfer, and these initiated by as few action potentials as necessary to transmit the minimum information needed for effectiveness. In fact, synapses save large amounts of energy by reducing the probability of transmitter release to a number far below 1 (discussed below). The cortex can be metabolically efficient by using synaptic plasticity to ensure that every transmitted signal delivers significant information to the postsynaptic neuron. An updated energy distribution approximates 60% for signaling and 40% for housekeeping, which includes processes such as lipid turnover, mitochondrial proton leak and actin treadmilling [Engl and Attwell 2015].

Counter-intuitively, Raichle and Gusnard [2002] point to the fact that the high rate of metabolism is essentially constant despite widely varying mental and motor activity. It is suggested that the brain's baseline use of energy is generally high, and a large amount of this metabolic activity is focused on maintaining a balance between excitatory and inhibitory activity by ongoing synaptic processes. They note that brain activation distinguishes itself from ongoing brain metabolism, where blood flow and glucose utilization increase faster than oxygen consumption. Coupling between cerebral blood flow (CBF) and cerebral metabolic rate of oxygen ($CMR_{O2}$) during cerebral activation is controversial [Valabrègue et al 2003], and are major determinants of the contrast in fMRI. Here, the modeling assumes a relaxation of the tight coupling between CBF and $CMR_{O2}$, allowing equations more flexibility to reflect experimental results.

A neuronal hypothesis suggests that astrocytes communicate neuronal energy demand to the vasculature in two ways, resulting in vasodilation by a feedforward signaling mechanism allowing for fast adaptation in anticipation of uneven energy requirements, as well as by metabolism-dependent mechanisms that react to energy requirements, but after the fact and therefore slower [Mergenthaler et al 2013, Howarth 2014].

While the metabolic rate of the brain is relatively constant, hard exercise does not significantly increase its overall fuel consumption, and sleep only reduces its metabolic rate by about 15% [Barros 2010]. However, the metabolic rate of individual neurons can vary significantly, requiring as much as multiples of 30-100 in fuel demand, depending on the neuron. The contrast between organ-level and cell-level demands might be attributable to a temporal averaging mechanism. It is suggested that the architecture of the tissue is such that active cells are surrounded by inactive cells that remain passive. Thus the global energy requirements can be managed at the nominal value. Based on a diffusion scale argument, Barros sets the maximum size of a metabolic unit in the brain tissue within which there is spatial averaging to be about 1 mm.

5.   Glucose Dynamics

A "bottom up" model of glucose dynamics in the brain is developed by Barros et al [2007] beginning with known microscopic properties of the GLUT transporters. Most of the brain's ATP turnover focuses on feeding neuronal $Na^+$ pumps during signaling. Cloutier et al [2009] suggest that cellular energetic requirements drive energy metabolism, given that many of the kinetic parameters of such metabolism are the least sensitive parameters in their model. They further provide support for the ANLS hypothesis. This is also proposed by Hitze et al [2010] where the brain acts "selfishly" via a cerebral demand mechanism to preserve brain mass even during inanition. Such a "brain-pull" mechanism is critical for survival given the high energy demands of the brain. The brain's need, supply, and demand for energy are increased by stress. The selfish brain theory is also mathematically modeled by Gödel et al [2010] in a five-part model.  These are: cerebral ATP, the energy in the body periphery (glucose, glycogen), blood glucose, insulin concentration as feedback signal of the brain, and ingestion regulation. This model consists of energy fluxes between compartments and signals directing energy fluxes, with the model providing a realistic qualitative energy metabolism behavior.

Jolivet et al [2009] introduce a "top-down" approach that utilizes datasets of average glucose and oxygen utilization in the brain at different activation states. Their analysis suggests that astrocytes take up a significant amount of glucose while primarily neurons consume oxygen. Such utilizations increase with increased activation, with astrocytes taking up more glucose than can be oxidized while neurons oxidizing more glucose-derived metabolites than the amount of glucose they take up, suggesting a transfer of glucose-derived metabolites from the astrocytes to the neurons, consistent with the astrocyte-neuron lactate shuttle (ANLS). Given that different cell types have different metabolic profiles, the above implies that there exist a broad and deep set of signaling mechanisms that regulate and coordinate these metabolic processes [Magistretti and Allaman 2015].

Barros and Deitmer [2010] continue the discussion and debate regarding the so-called astrocyte-neuron lactate shuttle. The mechanism put forward is that astrocytes take up glucose, transform much of it to lactate, which is then exported into neurons to be oxidized. They point to evidence that lactate acts as a signaling molecule involved in $Na^+$ sensing and in glucosensing, and it also couples metabolic activity to the modulation of vascular tone in the brain. But it also appears that it is not possible to definitively state the source cell or the sink cell for lactate, and perhaps even for glucose, since both are exchanged between adjacent regions of the neuropil via the interstitial space. Neurons are better suited to endure fuel shortages than other cells, including astrocytes, can react to their own increases in demand, and are better equipped to take up fuels than to release them. Neuron glucose transport capacity is about 12 times that of astrocytes, although this does not necessarily correlate with high metabolic rate. A conclusion is that there may not be "typical" neurons or astrocytes. Rather, there is a spectral range of such cells, implying that energy problems are solved locally via a spectrum of strategies. Support is provided by Mächler et al [2016] who point to *in vivo* evidence of a lactate gradient from astrocytes to neurons, which is a prerequisite of a lactate shuttle in that direction. And yet there is debate since there are several lines of evidence that neuron synapses rely on local glycolysis, that is, glycolytic enzymes, to rapidly metabolize glucose to provide fuel for key steps in synaptic vesicle recycling during synaptic activity [Ashrafi and Ryan 2017]. ANLS might be a backup during periods of reduced carbon supply. An understanding of presynaptic and axonal metabolism is critical, as it appears clear that dysregulation of neuronal energetics is linked to neurodegeneration.

While most energetics studies of brain matter have focused on grey matter, Harris and Attwell [2012] have calculated the energetics of white matter. White matter (~50%+ of brain volume) comprises mainly of myelinated axons, while grey matter (~50%- of brain volume) few myelinated axons. Grey matter contains most of the neuron somas while white matter is made up mostly of myelinated axons. Grey matter is made up of nerve cell bodies, and white matter is made up of fibers. Neurons of white matter have extended axons while grey matter neurons do not. It was found that the primary reason that white matter uses less energy than grey matter is that there are very few synapses, not due to the presence of energetically efficient myelinated axons. Additionally, the oligodendrocytes that create the myelin sheath for the axons require ATP to maintain their resting potential, are an addition to the energy budget, and that energetic cost is greater than the savings accrued due to the myelinated sheaths. In summary, the energy use of white matter is 3-fold lower than gray matter mainly because it has an 80-fold lower density of synapses.

Synaptic energy use, especially the requisite return to homeostasis after synaptic transmission, is a critical component of understanding brain energetic dysfunction and how these lead to neuropathology [Harris et al 2012]. There are a number of interesting and counterintuitive results. A cortical neuron receives about 8,000 synapses on its dendritic tree on average, and cannot possibly transmit all this information through its output axon. Thus, as a way to minimize energy use, there are an optimal number of synaptic transmissions, implying a low release probability, as mentioned earlier. There is an inverse relationship between release probability and release sites. The failure of synaptic transmission reduces this energy waste. The high energy demands at pre- and post-synaptic junctions places a large number of mitochondria and ATP production at these sites. Any disruptions to mitochondrial trafficking or function will definitely affect synaptic functioning, and lead to neurodegenerative diseases. An example, one of many, is the sensitivity of dopamine neurons to degeneration in Parkinson's disease [Pissadaki and Bolam 2013], possibly due to their orders of magnitude larger unmyelinated axonal arbor. Due to the resulting very high energy demand, such neurons are at risk of cell death when any stressor causes an excess in energy demand over supply.

Glutamate is considered to be the most important neurotransmitter in mammals. Its transport not only coordinates excitatory signaling needed for essentially all motor, sensory, and autonomic processing, but also plays a pivotal role in regulating brain energetics [Robinson and Jackson 2016]. Glutamate transport is both dependent on mitochondrial function, and is a regulator of mitochondrial function. It has been demonstrated that mitochondria are present in fine astrocytic processes, and they redistribute to provide a local source of ATP. Between 15% and 30% are mobile. A deep understanding of the coupling between brain energetics, mitochondria, and astrocytes can provide opportunities for therapeutic interventions in those instances when dysfunction occurs on any level. Neurometabolic couplings, mechanisms where astrocytes adjust energy production to meet neuronal energy needs, are particularly important when neurons undergo rapid changes in their firing rates, as during sleep/wake transitions [Petit and Magistretti 2016]. Interestingly, brain energetic expenditures during non-rapid eye movement sleep decreases only to about 85% of the waking value [DiNuzzo and Nedergaard 2017], confirming our prior discussion that the metabolic rate of the brain as a whole remains relatively constant whether the body is in hard exercise or in sleep. Even though the purpose of sleep is not fully understood, given the adverse effects of sleep deprivation, it is clearly needed. Important effects include plasticity of neuronal networks, homogenization of neuronal spiking rates, and clearance of brain lactate through the glymphatic system.

Glucose, glycogen and lactate are viewed as the metabolic trinity that couples astrocytes and neurons, not only ensuring functional brain energetics but also ATP turnover, astrocytic signaling, memory consolidation, and gene expression. The ANLS hypothesis is disputed, since how lactate is generated and used is a function of activity level. That is, exercise drives lactate into all brain cells, becoming an oxidative fuel, and spares glucose for cells, such as neurons, that almost exclusively rely on that source. Disruption of transport or metabolism of any of the triad results in changes in the activities of the other two [Dienel 2017]. Furthermore, there is a broader understanding of neuronal metabolic versatility, where brain cells can use fatty acids and ketone bodies for energy metabolism, as well as glucose and lactate. Neurons are completely capable of glucose uptake and glycolysis to sustain brain activity during brain stimulation, without any need for astrocytes or astrocyte-derived metabolites [Tang 2018]. Acetate, an alcohol by-product formed in the liver, can also be an alternative fuel for the brain, but can only be metabolized by astrocytes, not neurons. Under normal resting conditions, acetate can provide approximately 10%-15% of the energy demand of brain astrocytes. Monocarboxylate transporters most likely mediate glia-axon metabolic interactions, and facilitate the functioning of lactate as a preferred energy metabolite to support axon function. Disruption of these transporters contributes to neurodegeneration [Jha and Morrison 2018].

While there has to be a constant flow of energy to the neurons and glial cells, environmental energetic stressors lead to adaptive strengthening at the cellular and molecular levels [Camandola and Mattson 2017]. Physical exercise and food deprivation lead to synapse strengthening, formation of new synapses, and the production of new neurons from stem cells at the cellular level, and neuron resistance to a variety of stressors at the molecular level, bolstered via the activation of transcription factors that results in the expression of proteins. These changes add resistance to various neurodegenerative diseases.

6.  Concluding Summary

Brain energetics are central to proper functioning at the cellular, subcellular and system-wide scales. An understanding of the mechanisms of each scale, and how these couple, is critical to the comprehensive understanding necessary for the development of effective clinical interventions with minimal unintended consequences.

# Mitochondria

1.  Morphology and Function

Mitochondria are intracellular organelles, with primary functions including the maintenance of energy homeostasis, cell integrity and survival [Simcox and Reeve 2016]. Mitochondria comprise approximately 20% or more of cell volume, attesting to their importance. In the aggregate they comprise approximately 10% of body weight. These organelles produce up to 95% of a eukaryotic cell's energy (ATP) through oxidative phosphorylation [Tzameli 2012] driven by an electrochemical gradient created by the respiratory chain. They are double-membrane organelles with tightly packed cristae inside the inner membrane, resulting in a significant increase in surface area (much in the same way as the folded cerebral cortex) over which oxidative phosphorylation and maintenance of the proton gradient can occur. Cristae biogenesis, regulated through ATP synthase, which is also responsible for the majority of cellular energy production, closely links mitochondrial morphology to energy demand [Simcox and Reeve 2016]. The human body consumes on average a quantity of ATP per day that approximates its body weight [Zick and Reichert 2011].

     Mitochondria exist in varying numbers dependent on cell type and may form an intracellular network of interconnecting organelles, extending throughout the cytosol and in close contact with the nucleus, the endoplasmic reticulum, the Golgi network, and the cytoskeleton [Benard et al 2007]. The mitochondrial network is also known as the mitochondrial reticulum. The word reticulum is defined as a net-like formation or a network. "Mitochondrion," when translated from the Greek means "thread grain." Mitochondria may exist as small isolated particles, or as extended filaments, networks or clusters connected via intermitochondrial junctions. Single-section electron microscopic pictures of the cells only showed a two-dimensional image of the mitochondria, with the third dimension being lost. Serial-section images recovered that third dimension, showing filamentous mitochondria, frequently linked into networks [Skulachev 2001]. Extended mitochondria, and electrically coupled mitochondrial clusters, can serve as a system of effective power transmission in the form of membrane potential between remote parts of the cell. To avoid local damage due to a discharge of many of the organelles within the cell, the mitochondrial contacts by which networked mitochondria are coupled can be switched off (and on) as needed. Interestingly, isolated mitochondria unite into extended mitochondrial systems when the cell encounters energy shortages. (This reminds us of how astrocyte networks control the extent of their connections via gap junctions.) In tissues composed of large cells with high energy demands, extended mitochondrial systems occupy much of the cell

volume. Under certain physiological conditions and pathological conditions, such as a depletion of mtDNA, mitochondrial filaments and networks can decouple into single mitochondria. Extended mitochondrial systems exist only when their energy coupling and transmitting machineries are functioning normally [Skulachev 2001]. Muscle fibers require a specialized spatial organization of the mitochondrial network [Vinogradskaya et al 2014].

There is a coupling between mitochondrial morphology (network organization) and its bioenergetic function. Alternate configurations interact with bioenergetic properties in a way that is not fully understood, but the relationship is bidirectional [Benard et al 2007]. Wai and Langer [2016] connect mitochondria morphology to their functioning. In particular, how fusion and fission shape morphology, and that disruption of these, for example unopposed fission or fusion, adversely impact cellular and organismal metabolism, leading to potentially devastating dysfunction. Fusion and fission are also denoted as mitochondrial dynamics.

Frequent cycles of fusion and fission [Pagliuso et al 2018] adapt the morphology of the mitochondrial compartment to the metabolic needs of the cell, and optimize the organelle's bioenergetic capacity. Mitochondrial fusion optimizes function by spreading metabolites, enzymes, and mitochondrial gene products throughout the mitochondrial compartment, and counteracts the accumulation of mutations during aging. Extensive adaptations of mitochondria to bioenergetic conditions also occur at the level of the inner membrane ultrastructure and the remodeling of mitochondria cristae [Westermann 2012].

Fused mitochondria permit optimal functioning in respiratory active cells due to the mixing of the matrix and the inner membrane, allowing close cooperation by the respiratory machinery. Fusion engages the entire mitochondrial compartment in respiration, maximizing ATP synthesis. A sudden need for metabolic energy may lead to the formation of hyperfused mitochondrial networks as a result of cell stress. Interestingly, such short-term stress exposure in starvation results in fusion that optimizes mitochondrial function. This also plays a beneficial role for the maintenance of bioenergetic capacity in the long term. In a complementary way, fission contributes to the maintenance of bioenergetic capacity by eliminating irreversibly damaged mitochondria by autophagy [Westermann 2012]. Furthermore, the activity of key proteins of mitochondrial dynamics is regulated at multiple levels, in direct response to the bioenergetic state of the mitochondria. Three members of the Dynamin family, which are GTPase enzymes, are key components of the fission and fusion machineries [van der Bliek et al 2013]. We still have a limited knowledge of the mitochondrial proteome, but expect it to be customized to location.

Mitochondrial dysfunction is recognized as a key aspect of both acute and chronic allostatic states, that is, states of stress from which homeostasis is attempted. Genetic and environmental factors lend variability to mitochondrial stress response function [Manoli et al 2007].

It is clear that limited models of abnormal mitochondrial dynamics are insufficient to explain phenotypic variability in symptoms; information, and mechanisms need to function across multiple levels of organization [Eisner 2018]. Required are dynamic models of mechanisms at the molecular and organelle levels over long time periods. Purely descriptive representations cannot accurately picture multivariate dynamics. As suggested elsewhere in this paper, it appears that physiological and pathological processes result from biochemical, morphological and mechanical dynamics at more than one scale.

The mitochondrial network changes its shape and distribution in the cell largely via the evolutionarily conserved activities of mitochondrial fission, fusion, motility, and tethering. The connectivity of the network is governed by the relative rates of mitochondrial fission and fusion, suggesting that the two processes are coordinated as a result of energetic needs. In complex polarized cells such as neurons, mitochondria must be actively transported and tethered to, and maintained, in active synaptic regions. Tethers are important for positioning mitochondria within the overall cell structure, and also relative to other organelles. There is evidence that the four conserved activities – fission, fusion, motility and tethering – are interdependent and that the dysfunction of one activity can have consequences on another. For example, attenuation of fission disrupts the transport of mitochondria to neuronal synapses, resulting in detrimental effects on cell function. Tethering defects can reduce fission rates. A full understanding, therefore, requires an understanding of each process individually, and all processes as a coupled system [Lackner 2014].

The term mitochondrial plasticity is used to describe their adaptation to neuronal energy states via changes in morphology (form), function and position. Since the primary site of neuronal energy consumption is at the synapse (pre and post), it is there that mitochondria congregate and adapt to local energy needs via feedforward and feedback regulatory mechanisms [Rossi and Pekkurnaz 2019]. Furthermore, at the molecular level, mitochondria and axonal cytoskeleton tracks (such as actin filaments and microtubules) regulate mitochondrial distribution and dynamics. A motor-adaptor complex exists on the mitochondrial surface that contains kinesin and dynein. This complex contains the proteins Miro and Milton and is responsible for much, although not necessarily all, mitochondrial movement [Schwarz 2003]. Multiple signaling pathways converge on this complex to tailor mitochondrial positioning [Rossi and Pekkurnaz 2019]. Perhaps via an optimization interpretation, choices are made among the multiple signals via an evolutionarily refined weighting mechanism, resulting in particular mitochondrial movement. Similarly, mitochondrial plasticity in response to immediate energy needs at synapses can also be interpreted to be an optimization of energy availability and use where it is most needed.

We suggest that such "choices" are in some sense optimal and evolutionarily conserved to assure that energy supplied matches demands. Where this does not occur, due to insults to the system or to system defects, such as in mtDNA, the optimal choice becomes one where dysfunction occurs, resulting in pathologies and neurodegenerative diseases. While data can help create a framework for understanding how the distinct signaling pathways are correlated in space and time to assure mitochondrial bioenergetics and distribution throughout the synapse, mathematical models that represent the optimal choices can be powerful tools for a systematic understanding.

There is substantial evidence that the mechanical aspects of cellular and mitochondrial dynamics, and organelle dynamics generally, are intimately coupled to numerous biochemical processes related to healthy cell functioning, and to pathological developments [Feng and Kornmann 2018]. For example, ion channels are mechanosensitive, proteins can be curvature sensing, and the cytoskeleton and plasma membrane are force sensing. Organelles are dynamic and move around the cytoplasm, resulting in physical contact with each other, allowing them to exchange metabolites and information. This intermingling appears to be an important feature for the proper functioning of eukaryotic cells. It is hypothesized [Feng and Kornmann 2018] that biological phenomena observed at organelle contact locations are, at least, partially attributable to mechanical stimulation, for example, reticulum-induced mitochondrial fission. Mechanosensitive signaling pathways are critical components of a complete understanding of cellular and intracellular mechanisms [Moeendarbary and Harris 2014, Petridou et al 2017].

An understanding of the two-way coupling between the mechanical properties of cells and how physical loads on cells lead to biological and chemical responses is critical, in particular for the development of clinical diagnostics as well as therapeutically successful interventions. It is also known that many chemicals can alter the mechanical properties of living cells [Lim et al 2006]. As such, certain cellular mechanical properties can be utilized as indicators of health.

Based on theoretical arguments [Hoitzing et al 2015] a number of hypotheses are suggested regarding mitochondrial networks and their dynamics: (i) selective mitophagy is not required for quality control because selective fusion is sufficient; (ii) increased connectivity may have nonlinear effects on the diffusion rates of proteins; and (iii) fused networks can act to dampen biochemical fluctuations, making them less susceptible to the effects of such fluctuations. The structure of mitochondrial networks is also modeled using network tools and percolation theory [Zamponi et al 2018], which is generally used to model the movement and filtering of fluids through porous materials.

The mitochondria connect with the endoplasmic reticulum (ER) (both are tubular organelles) where proteins situated on opposing membrane faces may interact and tether the two organelles to each other. The ER-mitochondria interface provides a platform for the regulation of different processes: coordination of calcium transfer, the regulation of mitochondrial fission, the regulation of inflammasome formation, and the provision of membranes for autophagy. The ER regulates mitochondrial dynamics, and morphological changes, thus affecting cell health. The ER and mitochondria communicate reciprocally, transmit danger signals that can trigger multiple, synergistic responses, including increasing the number of ER-mitochondrial contact sites to allow for enhanced molecular transfers [Giorgi et al 2009, Marchi et al 2014]. Given the criticality of proper mitochondrial function to good health, an understanding of the ER-mitochondria interface is relevant to understanding human diseases that are believed to arise from mitochondrial dysfunction.

In addition to the ER, mitochondria have bidirectional communications with other cellular organelles, in particular, lysosomes, and peroxisomes. This communication is involved in the pathology of mitochondrial diseases [Diogo et al 2018]. Genetic defects in mitochondrial proteins cause a large group of diseases known as mitochondrial diseases, discussed below. One result of these diseases is that lysosomes and peroxisomes are affected structurally and functionally. In the other direction: many lysosomal and peroxisomal diseases perturb mitochondria; lysosomal storage diseases perturb peroxisomal metabolism and mitochondrial function; peroxisomal diseases often lead to perturbations of mitochondrial structure, redox balance and metabolism; saturation of lysosomal capacity is often observed in mitochondrial diseases, with accumulation of dysfunctional lysosomes and autophagosomes. Clearly, a full understanding of cellular metabolism and mitochondrial diseases requires an understanding of how the mitochondria communicates with the rest of the cell. While some understanding can be gathered by a reductionist consideration of the mitochondria, this can only be considered as a first step.

Mitochondrial motility, or trafficking, is critical for the survival of all cells, but likely more for neurons than all other cells, primarily due to exceptional neuronal morphology. Neurons extend their axons and dendrites distances of up to a meter, between two and three orders of magnitude greater than most other cells. The trafficking of mitochondria in neuronal axons, which lie flat and are typically about a micrometer in diameter, is along linear arrays of uniformly polarized microtubules, where the negative ends of the tubules are anchored in the cell body and the positive charge ends in the distal tips. Rather than forming a complex reticulum, in the axons mitochondria have separated from the reticulum and exist as discrete organelles of dimension typically 1-3 microns (micrometers), with those in dendrites tending to be longer. Mitochondria must provide the energy required for the proper functioning of the neurons, especially at the synapses, where energy demand is greatest. The mitochondria must traffic to high energy demand sites as needed, where such demand is a dynamic process and can change rapidly. Interestingly, about 10%-40% of the mitochondria are moving at any instant of time, and of those about half are moving away from the cell body (anterograde, kinesin-dependent) and half toward the cell body (retrograde, dynein-dependent). It is also speculated that the mechanism of movement and fusion are mechanistically coupled. A rich and complex regulatory machinery has evolved to manage the distribution of mitochondria over long distances and matching their distribution to a very dynamic demand for energy [Schwarz 2013].

Molecular machinery underlies mitochondrial movement and dynamics (quality control via fusion and fission). Mitochondria must navigate the highly branched paths of the complex neuron geometry, and know where and when to stop. A diverse and sophisticated molecular system is used by neurons to transport and anchor mitochondria. Machinery for fission and fusion can intersect with machinery for motility. Similarly, molecular machinery controls mitochondrial quality, using feedforward and feedback mechanisms. If there is molecular misregulation of mitochondria, neurodegeneration is possible. Such mitochondrial dysfunction, in conjunction with other malfunctions, can result in pathological conditions, or can be the direct cause of neurodegeneration [Vanhauwaert et al 2019].

In addition to their role in ATP production via oxidative phosphorylation, mitochondrial functions include iron-sulfur cluster formation, calcium handling, apoptosis and quality control through fission and fusion, cell signaling, cell repair and maintenance, and ROS production [Simcox and Reeve 2016]. ROS emission is a dynamic balance between production and scavenging [Kembro et al 2013]. As part of their critical metabolic functions, mitochondria also perform critical apoptosis functions [Vakifahmetoglu-Norberg et al 2017]. Mitochondrial dysfunction can arise from a host of causes, from genetic defects to other intra- and extracellular environmental instigators, and are a nexus pathological feature for incurable neurodegeneration [Correia and Moreira 2018], leading to motor, behavioral and cognitive loss of function, leading to death. Neuronal fate is tied to successful coupling between mitochondrial bioenergetics and dynamics (morphological changes, changes in transport and turnover). The importance of mitochondria in neuronal function is evidenced by their large numbers at synaptic locations.

It is clear that the mitochondria are a fundamental component for healthy living. In particular, their optimal functioning and efficient energy usage are necessary to avoid numerous pathologies. Optimalities are maintained at all levels of functioning, for example, preserving an optimal pool of mitochondrial nicotinamide adenine dinucleotide (NAD) [Tzameli 2012]. The stages of metabolism have been evolutionarily optimized and have become efficient.

The tight coupling between cellular bioenergetics, metabolism, the inner membrane structure, and mitochondrial function, as well as neuronal energy requirements at a high and continuous level, suggest that understanding these connections in a fundamental way is a link to understanding neurodegenerative diseases, bioenergetic dysfunction, and mitochondrial diseases, and hopefully to the identification and the enacting of clinical efforts for recovery.

2. Mitochondrial Disease and Neurodegenerative Disorders

Primary defects of mitochondrial function have been described for the five complexes of the respiratory chain, mtDNA replication and translation, respiratory chain cofactor synthesis, protein/solute import, and membrane structure and function. Secondary mitochondrial dysfunction relates to the impact of primary defects on diseases such as Alzheimer's, Huntington's, cancer, and in the aging process [Lemonde and Rahman 2014]. Some of these are discussed subsequently. Mitochondria also appear to be involved in the pathogenesis of multiple sclerosis [Adiele and Adiele 2019].

Mitochondrial dysfunction is initiated by glutamate excitotoxicity [Greenwood et al 2007], which may be implicated in the secondary cascades of traumatic brain injury, as well as in neurodegenerative diseases of the central nervous system.

Even relatively minor mitochondrial dysfunction leads to devastating systemic neurological and neurodegenerative disorders (APHALS), for example, Parkinson's disease, and Huntington's disease, which also has psychiatric components [Buhlman 2016]. Mitochondrial bioenergetics and dysfunction can lead to a failing heart [Sheeran and Pepe 2017]. ADP and Ca$^{++}$ signals contribute to the matching of energy supply and demand during changes in heart workload, and oscillations of mitochondrial ATP regulate the cardiac action potential [Wei et al 2011]. While many mitochondrial disorders are multisystemic, some are very specific, for example, optic neuropathy, sensorineural deafness, and type 2 diabetes mellitus [Schapira 2012].

Mitochondrial disease can be described as a group of diseases characterized by even small misfiring of mitochondrial energy production, usually in the respiratory chain within the five enzymatic complexes, which are housed in the so-called matrix. This respiratory chain is controlled by the nuclear DNA, nDNA, and the mitochondrial DNA, mtDNA. Due to the nature of mitochondrial genetics, that each cell has hundreds or thousands of mtDNA copies, during cell division genetic mutations in mtDNA are distributed to the daughter cells randomly. Daughter cells may have both mutant and wild (non-mutant) type mtDNA copies. With time, and further cell divisions, the tendency is for an accumulation of mutant cells. When the ratio of mutant to wild type mtDNA copies grows beyond a certain threshold, clinical symptoms of disease can occur. But with mutation, the phenotype may change, and different patients may manifest their diseases at different stages in different ways. The threshold is not a fixed number, but is random within a range. This randomness is evidenced by heterogeneous manifestations of mitochondrial disease [Kurt and Topal 2013].

Mitochondrial fission and fusion are processes by which the organelle maintains quality control. An imbalance between these, which are mediated by the action of GTPases, results in mitochondrial dysfunction [Panchal and Tiwari 2019]. Fusion is the means by which mitochondria exchange intermembrane and matrix contents, including mtDNA. It dilutes damaged mtDNA and lowers the threshold ratio between mutant and wild types. Fission provides a mechanism to isolate for elimination components that become damaged due to age or due to increased oxidative stress.

Mitochondrial dysfunction and defective dynamics are proposed as a key mechanism in the early stages of AD, with the cause given as energy loss. In particular, there is a direct proportion between mitochondrial oxidative phosphorylation in the AD brain and respective clinical disability. Such metabolic dysfunction correlates with abnormal mitochondrial morphology and fragmentation, and altered expression of mitochondrial fusion and fission proteins. In ALS, impaired mitochondrial transport in axons is observed, as well as fusion and fission abnormalities. HD is also connected to abnormal mitochondrial morphology. In PD, there is excessive fission, resulting in mitochondrial fragmentation. In these big four neurodegenerative diseases, abnormal morphology is linked with genetic and toxin causes. In connection with this, reactive astrocytosis and microglial activation play a role in inflammation, in part due to mitochondrial dysfunction. Common phenotypes are mitochondrial structural and functional defects [Joshi and Mochly-Rosen 2018]. In genetic forms of PD, disruptions in synaptic vesicle endocytosis (SVE) significantly contribute to disease pathogenesis. Mitochondria provide ATP to power SVE, which replenishes synaptic vesicles to sustain repeated release of neurotransmitters. Mitochondrial and lysosomal dysfunction have been found to be key cellular processes that contribute to PD pathogenesis in such genetic forms of PD [Nguyen et al 2019].

It is interesting to note and observe that mitochondrial dysfunction can be a cause of age-related neurodegenerative diseases, but can also be implicated in the progress of such diseases, their pathogenesis. It is important to recognize that there are numerous progression paths and possible initiations of neurodegenerative diseases, and mitochondrial dysfunction can be an initiator or a contributor, or both in the case where the disease initiates feedback loops to the mitochondria, creating a spiral of degeneration. Mitochondrial dysfunction can precede or occur concurrently with an onset of pathologies.

Aging can lead to an accumulation of reactive oxygen species (ROS), which are major contributors to oxidative stress, due to the imbalance between the production of ROS and their oxidation, which then affects mitochondrial respiratory chain function. Membrane permeability is altered as is calcium homeostasis, increasing heteroplasmic mtDNA and weakening mitochondrial defense systems. It is generally agreed that mtDNA mutation, oxidative damage (which can be minimized via exercise and caloric restriction), and/or aggregation of mitochondrial proteins leading to abnormal mitochondrial morphology are the main reasons for the organelle's dysfunction. Cumulative oxidative stress is viewed as the critical factor that precedes a cascade to dysfunction [Elfawy and Das 2019].

There is evidence that mitochondrial dysfunction is linked to Rett Syndrome, and Autism Spectrum Disorders (ASD). In the latter case, links between abnormalities in mtDNA and the pathogenesis of ASD, which is a dynamic system of metabolic and immune anomalies involving many organ systems, including the brain, have diverse clinical manifestations and many genetic and environmental factors implicated in its development [Castora 2019]. Mitochondrial dysfunction may be present in up to 80% of children with ASD. Given the central role of mitochondria in energetics and the immune system, the high correlation between ASD and mitochondrial dysfunction is understandable. Additionally, one quarter of patients with mitochondrial disease experience epilepsy. There is strong evidence that metabolic change and mitochondrial dysfunction could be an important pathogenic process in epileptogenesis [Chan et al 2019]. These can be due to respiratory chain deficiency in both astrocytes and interneurons.

Due to the pleiotropic nature of nearly all mediators involved in mitochondrial function, quality control and cell death execution, interventions that target a single protein can have unexpected large-scale effects in multiple pathways, challenging those planning interventions [Sedlackova and Korolchuk 2019].

Mitochondrial dysfunction and disease are at the core of the initiation and progression of the major neurodegenerative and neuropsychiatric diseases. A key to understanding the disease pathology requires quantitative models of energetics at more than one scale, something that is currently lacking.

Mitochondrial diseases have a major relevance to military health. Approximately one third of the 1990-1991 Gulf War veterans – in the range of 175,000-250,000 soldiers – developed chronic multisymptom health problems with mechanisms that adversely affect mitochondria. Symptoms include muscle and brain fatigue with variable additional domains affected. Such chronic and long-lasting (to this day) multisymptom health problems are known by the term "Gulf War Illness" (GWI). In addition to the above, there are elevated instances of cognitive difficulties, muscle pain and weakness, shortness of breath, gastrointestinal problems, sleep problems, and behavioral changes [Koslik et al 2014].

Environmental factors are clearly implicated in GWI. Exposures to excessive and unique environments, as well as combinations of toxic exposures, particularly acetylcholinesterase inhibitor (AChEi), have led to toxic and lethal results via ROS and mitochondrial dysfunction (MD). These two outcomes are tightly coupled due to the fact that mitochondria are both target and source of ROS. It has been shown [Koslik et al 2014] that the post-exercise phosphocreatine-recovery time constant (PCr-R), a marker that serves as an estimate of net oxidative ATP synthesis, is a robust index of MD. A comparison of this index for GW veterans (GWV) to a control group shows that values of PCr-R for veterans can be more than twice those of the control group. Elevated rates of amyotrophic lateral sclerosis (also linked with MD) are also observed in GWV.

Elevated levels of mtDNA content and damage in circulation are markers for MD [Chen et al 2017]. Symptoms in GWV to control groups are reported in the following approximate ratios: fatigue 4/1; pain 21/1; neurological/cognitive mood 5/1; skin symptoms 2/0; gastrointestinal 4/0; respiratory 2/0. It was found that mtDNA damage in GWV is 20% greater than in the control group. Levels of nuclear DNA lesions were also elevated in GWI. All these results taken together are interpreted as evidence that mitochondrial dysfunction is involved in the pathobiology of GWI, and helps explain the persistence of the illness over 25 years, but not universally, as certain mitochondrial haplogroups are known to offer protection for specific neurodegenerative diseases [Chen et al 2017].

3.  Modeling

Given the significance of mitochondrial function and dysfunction, there have been a small number of works on the mathematical modeling of various aspects of their functions [Magnus and Keizer 1997, 1998a,b]. These models can be quite complex, with simpler derivative models attempted [Bertram et al 2006, Saa and Siqueira 2013]. These quantitative models governed ATP production during glucose metabolism via oxidative phosphorylation. The ability to connect mitochondrial dynamics mathematically to neuronal spike generation can be a powerful tool in disease modeling and enable clinical interventions that are not yet possible otherwise [Venkateswaran et al 2012]. The challenge is to populate the equation parameters, currently not fully possible.

The goal of computational models is to provide a general mathematical framework for the description of physiological system behavior under broad conditions. For example, an understanding of cellular function and dysfunction relies on a quantitative description of mitochondrial function and dysfunction. The larger framework requires computational models that span multiple scales: molecular, cellular, tissue, and whole-organ self-consistent models that naturally integrate across disparate scales. These mathematical models must satisfy the laws of chemical (bio)physics, of course, rather than be just products of data fitting. While data-driven models can accurately represent a narrow slice of reality, they do not accurately extrapolate beyond the originating data, and do not integrate across scales, much less provide a first-principles understanding. Data-driven models at different scales do not match at their boundaries. A computational model that meets the above attributes [Beard 2005] for the mitochondrial respiratory chain is derived to appropriately balance mass, charge, and free energy transduction. The components of this validated model are: the reactions at complexes I, III, and IV of the electron transport system, ATP synthesis at $F_1F_0$ ATPase, substrate transporters including adenine nucleotide translocase and the phosphate-hydrogen co-transporter, and cation fluxes across the inner membrane including fluxes through the $K^+/H^+$ antiporter and passive $H^+$ and $K^+$ permeation.

4. Concluding Summary

Given the central role played by mitochondria in cellular energetics, and that mitochondria are involved in almost all neurological disorders, to understand how mitochondria function is at the core of understanding neurological health, and also understand how dysfunction results in many of the most debilitating and fatal neurodegenerative diseases. This understanding can lead to therapeutic interventions to improve the health of many millions of people who suffer from these diseases.

The focus of our research efforts is an understanding of how and why mitochondria move, stop and anchor, undergo fusion or fission, degrade on the spatial and temporal scales needed to meet the broad spectrum of energy needs, as well as $Ca^{++}$ buffering needs, throughout the neuron. Such complexity suggests the existence of energy-based mechanisms that operate locally (sub-optimizations) and globally across multiple cells and beyond (full optimizations).

# Traumatic Brain Injury

1. Introduction

Numerous and significant advances have been made to our understanding of the human brain, how it functions on multiple scales, and how its various processes function biochemically and biophysically. While numerous challenges exist, real progress has been made, especially in the past several decades. One important clinical goal is the translation of this newfound knowledge and understanding to traumatic brain injury (TBI) rehabilitation.

Just as the brain operates on numerous scales, TBI, from initiation to reactions, is a multiscale event that evolves over timescales from the order of the initiating insult (μs) to days-weeks-months-years. TBI is complex and heterogeneous, with dynamics that are very sensitive to the initial conditions (location and morphological variations, insult object velocities and accelerations) and the health and genetic predispositions of the injured person. In the field of nonlinear dynamics, such sensitivities result in a spectrum of possible dynamics, with some paths leading to instabilities, and others to equilibria or homeostasis. Sometimes, we are more interested in the stability characteristics of the ensemble of TBI system dynamics rather than the particular trajectory since ensemble results allow us to determine the range of system parameter values that assure stability, or instability. Other times we seek a sample TBI response from the ensemble in order to observe a time history, perhaps to assess a particular clinical intervention.

In the next section, we discuss glial cells, in particular astrocytes and NG2-glia, due to their fundamental importance to TBI response. While other glial cells are also fundamental to this response, we focus on astrocytes and NG2-glia in order to suggest a possible approach to the modeling, which can be extended to other glia, such as oligodendrocytes and radial glia.

To understand TBI, we need to understand the fundamental aspects of the interplay and coupling between astrocytes and the synapses during normal functioning, and the biochemical processes that are initiated by microglia that drive astrocytes into a neurotoxic state. In a similar way, the dynamics of pre-TBI NG2-glia and their evolution post TBI can suggest clinical interventions. We note that the various glial subtypes behave differently prior to injury and post injury. For example, astrocytes react and function differently at early stages after injury than at later stages, in fundamental ways, with significant implications for the design of therapeutic measures as well as for the patients.

The dynamic map that is being created by researchers of glial types and functions is not only time-dependent, but also heterogeneous, and, with regard to TBI, these functions are sensitive to the characteristics of the originating insult and its location. Following TBI, astrocytes undergo a diverse and context-specific set of rapid changes in gene expression, morphology and function, collectively referred to as astrocyte reactivity [Dimou and Gallo 2015, Gotz et al 2015, Burda and Sofroniew 2017]. There exist a molecularly diverse set of subtypes of astrocytes that may serve a multitude of beneficial or deleterious roles. A key focus for TBI research is to understand the key molecular and cellular mechanisms driving such responses, and the functional implications of different forms of astrocyte reactivity [Burda and Sofroniew 2017].

2. Background and Significance

Traumatic Brain Injury (TBI) is a major cause of disability and death worldwide. Accidents and traumas acquired in war result in devastating injuries that, even if survived, result in severely limited lives of the injured as well as their supportive families. Various TBI classifications exist, the commonly used Glasgow Coma Index, or by use of pathological features. TBI damage may be classified to be focal or diffuse, the latter sometimes manifesting little obvious initial damage without the use of sophisticated imaging techniques, but potentially deadly even over relatively long time scales.

TBI includes concussion, contusion (including coup contrecoup), diffuse axonal injury (DAI), traumatic subarachnoid hemorrhage, and hematoma [Al-Sarraj 2016]. It is common that TBI results in more than one of these types of injuries. Secondary injuries occur as a result of the body's inflammatory response to the primary injury. This inflammation within the rigid skull of fixed volume results in an increase in pressure within the head. How to delineate and define TBI has been fundamental for research advances, allowing groups to compare and interpret results on an identical basis of understanding [Menon et al 2010]. As we discuss below, other secondary effects occur as part of a complex sequence of biochemical reactions that are not yet fully understood.

Widespread axonal damage, known as diffuse axonal injury (DAI), not necessarily immediately debilitating or fatal, can trigger secondary cellular processes and biochemical reactions that create unstable inflammation, resulting in greater damage than was caused by the primary injury. DAI is a shearing and stretching of nerve cells due to a rapid acceleration/deceleration motion of the brain. DAI is the most prevalent pathology of coup contrecoup. There is evidence that a single TBI can be a cause of later onset of neurodegenerative disorders such as Alzheimer's disease. Additional pathways to clinical dysfunction appear to exist, even from morphologically intact axons with disrupted physiology. An understanding of these pathways provides important therapeutic targets in the treatment of TBI, with mitigation of neurodegeneration [Johnson et al 2013, Rishal and Fainzilber 2014].

Substantial efforts have been made to understand the mechanical aspects of impact and how such forces translate into the brain at large and small dimensional scales [Hardy et al 1994, Hemphill et al 2015, LaPlaca et al 2007]. Advanced constitutive models are used with the goal of relating mechanical damage to physiological brain dysfunction [El Sayed et al 2008].

A significant challenge is to ascertain the mechanical properties of the brain, at all these scales, upon which the biomechanical models are based [Chatelin et al 2010, Jin et al 2013, de Rooij and Kuhl 2016, Zhao et al 2018].
TBI mortality rates range from 20% to 50%, depending on the initiating trauma, and these percentages have been declining recently, especially in wealthier societies where the emergency infrastructures are more advanced. In the US alone there are an estimated 2.1 million TBI cases per year, causing about 100,000 deaths per year and 500,000 hospitalizations [Büki and Povlishock 2006]. A broad spectrum of initiation and response mechanisms exist for TBI, but due to the hidden nature of some of these injuries, and the variety of individual reactions, treatment decisions challenge those at the front lines [Peebles and Cruz 2018]. Yet there are fundamental neurochemical and metabolic responses to TBI that suggest therapeutic approaches for early intervention [Prins et al 2013]. Using lactate as a fuel after TBI has been proposed, but the astrocyte-neuron lactate shuttle, which is at least controversial to some and opposed by others, needs to be carefully evaluated before being utilized in patient care [Dienel 2014]. This debate on the shuttle has been discussed earlier in this paper.

TBI has been viewed as primarily a male problem, due to the involvement of more men in activities that increase the risk of TBI, and resulting in using male animals for preclinical research. However, this view has evolved as the population ages and females are more involved in activities that can lead to TBI. Thus, preclinical efforts are beginning to account for sex-related responses [Spani et al 2018].

While initial diagnosis is challenging, and early treatment – within about an hour – is critical, there are treatments for the deteriorating pathological stages encountered. Medications for secondary injuries are still in the research stage, and least understood. This is true even though there have been tremendous advances in understanding neuro-glia systems, their biochemical couplings, and the genetic foundations of how the brain both recovers and deteriorates after TBI.

The primary approaches to understanding the biomechanics of TBI, as well as the pathophysiology of TBI, require *in vivo* and *in vitro* experiments, tests, and examinations of those who have succumbed to their injuries. The brain responds to TBI with complex, interdependent processes, with multidimensional dynamics, in a cascading way [Prins et al 2003, Bernick et al 2011].

Animal models have provided the community with the overwhelming majority of data and subsequent understanding, translating that understanding clinically [Laurer et al 2000]. A smaller component of our understanding is derived from theoretical (mathematical) and computational models. Often there are not enough data to populate the theoretical models. But the derivations of such models open up the field to broad development, suggesting new avenues of research via *in silico* experiments.

3. Cellular Level Behavior

Given the widespread and devastating aspects of diffuse axonal injury, we focus here on DAI, how glial cells respond, in particular the dynamics of astrocytic and NG2-glia behavior and response post TBI and over long time scales. A recent symposium broadly focused on glia-neuron interactions [Glia-Neuron 2018].

Here we discuss damage and response at the neuronal and glial level. An understanding at this scale requires attention beyond the neuron, especially for the development of rehabilitation and regeneration treatments to be applied at this biochemical scale.

The secondary effects discussed above can progress for years after TBI. One subgroup of TBI, DAI, results in traumatic axonal injury (TAI), a major driver of mortality and functional impairment [Hill et al 2016]. DAI pathology includes secondary physiological damage, for example, twisting and misalignment of axon microtubules and separation, resulting in loss of axonal transport [Kallakuri et al 2012], progressive swelling and degeneration. Secondary injuries result from biochemical cascades that occur in response to primary injuries. Axonal primary mechanical damage leads to intra-axonal calcium influx, resulting in destruction of the cytoskeleton, disrupting axoplasmic transport, and ischemia [Blumbergs 1998]. The $Ca^{++}$ influx activates cysteine proteases, calpain and caspase, which play a pivotal role in cytoskeleton destruction. This pathological progression also initiates mitochondrial injury, resulting in the release of cytochrome-c. The resulting degradation of the axonal cytoskeleton causes local axonal failure and disconnection [Büki and Povlishock 2006]. $Ca^{++}$ accumulation also produces reactive oxygen species (ROS), which are implicated in the pathology of TBI through the oxidative stress mechanism, and can also promote inner mitochondrial membrane permeability that forces the rupture of the outer membrane and the release of cytochrome-c that results in cell death through a chain of events [Gupta and Sen 2016]. Pathological progression to a particular end is not unique.

Variations in post-traumatic cellular bioenergetics and mitochondrial function may trigger or accelerate these damaging secondary intracellular cascades that follow the primary injury, resulting in delayed secondary manifestations of injury and inflammatory responses [Watson et al 2014].

This coupling between TBI and metabolic dysfunction due to mitochondrial damage can result in quick cell death due to energy failure. Diet (omega-3 fatty acids) can modulate the interaction between mitochondrial homeostasis and synaptic plasticity as a mechanism that can regulate the capacity of the brain to resist TBI [Agrawal et al 2014]. A better understanding of TBI from the clinical perspective for therapeutic strategies clearly requires an understanding of mitochondrial functioning and coupling with the metabolic aspects of neuronal and glial functioning.

TBI-induced cellular-event cascades include potassium efflux, $Ca^{++}$ accumulation, glutamate release, and increased oxidative stress that contribute to reduced ATP production with an increase in cerebral metabolic rates for glucose that can last for days and is a function of TBI severity. Reduced ATP production in conjunction with secondary insults result in secondary cellular damage. Increased energy demand, or a reduction in the availability of glucose after TBI, eventually reduces the functioning and survivability of neurons [Bartnik-Olson 2012]. As might be expected then, mitochondrial dysfunction appears to play a key role in the pathophysiology of TBI, in particular in reductions in mitochondrial state 3 respiratory rates (ADP-stimulated respiration) due to an uncoupling of the mitochondrial electron transport chain and mitochondrial permeability transition. The resultant energy loss contributes to cell death.

An additional theory proposes that DAI is triggered by myelin debris from injured axons that cause inflammation as well as expression of cytokines, resulting in a cascading DAI [Yang et al 2011].

An understanding of how axons elongate in response to forces is needed in order to develop neurological treatment strategies. Under normal conditions, such forces result in normal axonal growth, known as passive stretching or towed growth. Mechanical signaling occurs and neurons respond by increasing protein synthesis and transport [Suter and Miller 2011]. This is still not fully understood. It has been suggested that neurofilament proteins, key structural cytoskeletal elements, may be important contributors to axonal tensile strength, and thus, to how axons respond to TBI. Insights on these can contribute to the development of therapeutic strategies, hopefully to reverse neuronal degeneration following TBI [Siedler et al 2014].

Clinical challenges in drug trials to ameliorate TBI (DAI/TAI) are likely due to findings that a spectrum of cellular injuries occurs, at different locations, involving different receptors and biochemical pathways. Mathematical models that are linear, homogeneous and isotropic will not represent the brain accurately. There are critical factors that differentiate impact types, directions, and head movements. Characteristics of acceleration/deceleration, longitudinal forces coupled with shear strain forces resulting from rotational accelerations/decelerations, result in head/brain motions that lead to inelastic and permanent damage and ruptures.

The secondary biochemical cascade of events can result in death to cells that have survived the primary injury, partly due to calcium influx into the cell. The therapeutic window may last from hours to days after injury, but this is vague and unknown in advance [Meythaler et al 2001], and is highly patient/circumstance dependent [Corbo and Tripathi 2004, Maurya et al 2006]. Diagnostic tools are critical to establishing the extent and nature of the damage, thus offering opportunities for interventions [Li and Feng 2009, Mohammadipour and Alemi 2017]. Biomarkers of axonal injury provide valuable guideposts for early detection of injuries that routine neuroimaging cannot detect [Manivannan et al 2018].

Axonal injury directly affects dendritic morphology, a critical byproduct of TAI, and such changes in dendritic structure may be the cause of cognitive decline after TBI. How TAI affects changes in dendritic structure is less understood. It has been observed that TAI alters dendrite morphology by causing periodic swelling, called dendritic beading, resulting in, it is believed, a disruption in ionic homeostasis. But given the link between such structures and cognition, an understanding is required for the design of clinical solutions [Monnerie et al 2010]. The formation of dendritic beads is also correlated to changes in mitochondrial morphology and function, with the majority of beads containing a dysfunctional mitochondrion [Greenwood et al 2007]. Alzheimer's disease is associated with dendritic beads.

Connections between TBI and Alzheimer's disease (AD) are made at the cerebrovascular level. It has been suggested that cerebrovascular pathology, which is a key element in both conditions, could indicate a mechanistic link between Aβ/tau deposition after TBI and the development of dementia and chronic traumatic encephalopathy [Ramos-Cejudo et al 2018]. Mitochondrial disease and dysfunction, already discussed, is linked to AD, both as a cause as well as an outcome of the disease. Further investigation of the connection between TBI and mitochondrial dysfunction at the vascular level, especially regarding Aβ/tau deposits typically found near degenerating vessels, is warranted.

While there are multiple scales of damage to the brain in response to biomechanical insult, it is at the cellular level that the damage occurs and the brain responds, and the loss and biochemical rehabilitation takes place. Axonal loss results in progressive neurodegeneration, and novel therapies have made minimal and inconclusive impact on clinical outcomes. More recent research outcomes resulting in a deeper understanding of the cellular and molecular mechanisms that are initiated due to injury hold the promise that such understanding can translate to therapies at that scale. We will discuss how glia, in particular astrocytes and NG2-glia, appear to play significant roles in complex and not fully understood ways in such mechanisms and therapies.

Multiscale modeling approaches to predicting axonal strains resulting from TBI continue to make progress. Finite element methods, however, are challenged when scales range from organ (10-20 cm) to tissue (1 cm) to bundle (100-300 μm) to axon (1-10 μm) [Wright and Ramesh 2012]. One benefit of multiscale models is that axonal strain can be used as a measure of TBI. Such a coupling between the cellular mechanisms of damage and the mechanical loading at the macroscale is needed. One significant conclusion from multiscale computations is that tissue strains do not scale with maximum axonal strain in general. This is due to axonal heterogeneities in orientation, stiffness property gradients between the axons and the local surrounding medium, and the anisotropic properties of the axons [Cloots et al 2013].

4. Glial Cells

"Virtually every aspect of brain development and function involves a neuron-glial partnership [Barres 2003]." Glial cells (astrocytes) perform transmitter cleanup functions, maintain extracellular ion levels, and perform active control functions on synaptogenesis, synapse number, function, and plasticity. They (oligodendrocytes and Schwann cells) provide neuronal insulation and trophic support, control axon diameter, and the clustering of ion channels. Schwann cells also control regeneration of axons and the function of synapses at the neuromuscular junction [Barres 2003, Chung and Allen 2015]. Radial cells generate neurons and later glial cells, and participate and initiate a host of critical expressions. NG2-glia are observed to react very fast with a very large increase in proliferation (100-fold baseline in 1-3 days) at injury sites [Dimou and Gotz 2014], with still unclear outcomes and impact. As a part of this partnership between neurons and glia, there is evidence that neurons provide trophic support for glia.

      Glial cells, or neuroglia, maintain homeostasis, form myelin, and provide protection and support for neurons in the central nervous system. One glial function related to TBI is the destruction of pathogens and removal of dead neurons. On average, glial cells make up about fifty percent of the cells in the human brain. Some differentiated glial cells have reactivated stem and progenitor properties after insult. An understanding of which subtypes offer potential positive roles in repair can impact clinical response to TBI [Dimou and Gotz 2014].

      Microglia, which comprise approximately 10-15% of all cells found in the brain, are the resident macrophage cells. They play a dual and critical role in the brain following TBI, initially promoting beneficial neurorestorative processes, but also producing inflammatory and cytotoxic mediators that hinder repair and cause cell death. An understanding of the mechanisms that control microglial shifts can suggest therapeutic strategies for TBI [Loane and Kumar 2016].

      One response to damage to the central nervous system is gliosis, a hypertrophy of glial cells, resulting in a complex and not well-understood series of biochemical cascades, initially for the repair and containment of the damage, and then for remyelination and regeneration of axons and dendrites. As part of this sequence of events, an abnormal increase of astrocytes, called astrogliosis, occurs with a change in astrocyte morphology and molecular expression, resulting in a scar, and sometimes inhibition of axon regeneration. There is still debate regarding the scar and its role in

TBI recovery, as well as regarding the time-dependent functions of the glia. (For example, while a scar does not form in wounded zebra fish, there is extensive regeneration of wounds.) Do the glia respond to damage and disease or do they make things worse, or both? The glia have a wide spectrum of functions, with cell-to-cell communications critical to the normal and efficient operations of the central nervous system. And yet, glial biology is not well understood.

Neuron-glia communications govern the formation of synapses and plasticity, while a breakdown in communications between the cells may be a significant cause of neural pathologies and common degenerative disorders [Shaham 2005, Stogsdill and Eroglu 2017]. Neuronal connections during development are often inappropriate, excessive or transient, requiring axonal pruning by glia, sometimes in a survival of the fittest competition. Experience-dependent synapse elimination, as a result of activity-dependent mechanisms, is critical for proper functioning, but the cellular and molecular mechanisms are only beginning to be understood [Corty and Freeman 2013, Allen and Eroglu 2017].

How individual glial cells respond to traumatic events is still under debate and study. From a morphological (structural) perspective, data is lacking since most research has focused on their electrophysiological and biochemical properties, and yet in order to understand and model brain reaction and response to traumatic injury, such properties are needed. Glial cells are softer than their neighboring neurons, and both types of cells are very soft, like rubber elastic [Lu et al 2006]. Glial cells can neither serve as structural support cells nor as glue, as originally thought.

Astroglia perform many functions, including biochemical support of the blood-brain barrier, maintenance of extracellular ion balance, and in response to traumatic injuries, with repair and scarring processes. There are several forms of astrocytes, located heterogeneously, each specialized to particular functions during specific times. They are active at synapses, secreting and absorbing neural transmitters, denoted as the tripartite synapse, referring to the pre and post synapse along with the astrocyte. Astrocytes are critical in health and disease/damage [Clarke and Barres 2013, Liddelow and Barres 2015, Verkhratsky and Nedergaard 2017].

Astrogliosis is a spectrum of reactions that are context-dependent, and can result in a continuum of progressive changes in gene expression and cellular changes. It is complex and heterogeneous. Reactive astrogliosis and glial scar formation appear to have important roles in determining clinical outcomes to central nervous system trauma. There is evidence that reactive astrogliosis with scar formation can inhibit axonal generation after such trauma, as well as exert essential neuroprotective functions [Sofroniew and Vinters 2010]. The existence of molecularly diverse subtypes of reactive astrocytes that serve a multitude of beneficial or deleterious roles confirms that this is a central question in the study of neurobiology of disease and injury [Burda and Sofroniew 2017]. TBI mechanopathogenesis is a trigger for astrocyte reactivity. This is how mechanical forces are translated into cellular dysfunction and damage. Understanding how this happens is fundamental to the development of therapeutic strategies for TBI, in particular, for milder forms of TBI that may cause widespread diffuse tissue damage but with little or no severe focal tissue damage [Burda et al 2016].

In response to sensing changes in neural activity and extracellular space composition, astrocytes exert homeostatic mechanisms critical for maintaining neural circuit function, such as buffering neurotransmitters, modulating extracellular osmolarity, and calibrating neurovascular coupling. In addition to upholding normal brain activities, astrocytes respond to diverse forms of brain injury with heterogeneous and progressive changes of gene expression, morphology, proliferative capacity and function that are collectively referred to as reactive astrogliosis. Destructive mechanical forces cause tissue damage in traumatic brain injury setting in motion complex events that disrupt central nervous system (CNS) homeostasis. This disruption triggers diverse multi-cellular responses that evolve over time and can lead either to neural repair or secondary cellular injury, depending on the severity of the insult. In response to TBI, astrocytes in different cellular microenvironments tune their reactivity to varying degrees of axonal injury, vascular disruption, ischemia and inflammation [Gotz et al 2015, Burda et al 2016].

More recently studied in depth, NG2-glia are widely dispersed throughout the adult brain and CNS, but the full spectrum of their function is still generally not known, but remyelination after injury appears to be one function. NG2-glia respond to injury and pathological conditions by transformations in cell morphology and proliferation rates. The specifics are sensitive to the nature of the insult and the age of the injured. The response is very fast, sometimes within one day, is highly dynamic, and the cells migrate toward the injury site. By the end of 28 days their numbers have decreased to their original density. Some hypothesize that the NG2-glia are responsible for wound closure and scar formation, to which axonal growth inhibition has been attributed. Positive and negative NG2-glia responses to pathological insults result in numerous explanatory hypotheses that may suggest new therapeutic strategies [Dimou and Gallo 2015].

There are selective signaling in the glia, in heterogeneous ways, very specific to insult type and location. The regulation of selective signaling might suggest opportunities for innovative therapeutic approaches to TBI [Dimou and Gotz 2014], in particular if an understanding can be developed regarding which glial subtypes create an adverse environment for the survival of new neurons, and thereby be potentially controllable.

5. Theoretical Modeling

Variational methods have been used for the derivation of theoretical neuronal or brain function models: nervous excitation [Dickel 1989], ionic solutions with applications to enzymes, ion channels and catalysis [Eisenberg 2011], energy coding the neural physical circuit [Chuankui 2012], and an electromechanical model of neuronal dynamics [Drapaca 2015].

Even though the human brain comprises about 2% of the body's mass, it requires about 20% of the body's oxygen and glucose to make the ATP it needs. The biochemical reactions and processes are governed by equations that dynamically balance various forms of energy. Of course, the laws of physics are satisfied at all scales and in all forms. Astrocytes play a major role by capturing nutrient molecules from the blood stream and converting them into the energy required by the neural synapses, and, uniquely, also storing the glucose from the blood stream in the form of glycogen. Other energy exchanges occur between neurons and astrocytes, for example, the retrieval of the neurotransmitter glutamate from neuronal synapses by astrocytes, a process that is energy expensive.

An understanding of astrocyte energy metabolism enhances our ability to model astrocyte-neuron interactions, including energy exchanges, as part of the glial processes that are critical to neuronal functioning [Hertz et al 2007, Prebil et al 2011]. NG2-glia react rapidly to any type of injury with strong potential to repopulate areas of lesion. That NG2-glia are found in the healthy as well as pathological human brain suggests a possible role for the development of strategies to promote myelin repair. The identification of factors that control their proliferation and differentiation may suggest why the adult central nervous system fails to remyelinate after TBI. Endogenous NG2-glia, therefore, represent a viable and effective target to regenerate oligodendrocytes and neurons [Dimou and Gallo 2015].

Normal functioning of astrocytes and NG2-glia are altered by TBI, as discussed. The possibility of deriving a set of equations that govern these energy exchanges, interactions, and storage during normal functioning, and then post-TBI can be very valuable to our understanding, and open clinical options for healing and regeneration. Of particular value for clinicians is the possibility of testing specific treatments as inputs to the theoretical models, and studying the models' estimates of clinical results.

Of interest are reduced order models of astrocytes as they react to insult, that is, astrogliosis [Kimelberg and Nedergaard 2010, Dimou and Gotz 2014, Burda et al 2016] as well as NG2-glia phenotype changes [Dimou and Gallo 2015]. As we have discussed, astrogliosis is both complex and heterogeneous, with emerging evidence that different forms of astrocytes function via different mechanisms post-injury in the following ways: (i) regulation of inflammation, (ii) isolation of lesions and protection of adjacent neural tissue, (iii) regulation of the blood-brain barrier, and (iv) synaptic plasticity and neural circuit reorganization [Burda et al 2016]. From Burda et al:

"Astrocytes sense and respond to mechanical strain after TBI. Physical strain deforms flexible networks of intermediate filaments within astrocytes and activates ion influx through mechanosensitive cation channels. Rises in intracellular calcium causes astrocyte ATP release that signals in an autocrine or paracrine manner, driving multiple intra- and inter-cellular signaling pathways, and inducing the release of endothelin-1,

MMP9 and glutamate. Depending on the severity of the mechanical insult, astrocyte reactivity may involve complex changes in phenotype and function that respond to and influence neuroinflammatory responses to injury as well as mechanisms of secondary TBI pathogenesis. Trauma also causes astrocytes to release GFAP and calcium-binding S100B that may serve as biomarkers of TBI severity." Each of these is an energy-related process, either energy release, energy creation, or energy dissipation. Initially, mechanical strain results in an ion flux. The rise in calcium results in ATP release, signaling the release of endothelin-1 (a protein), MMP9 (an enzyme), and glutamate (an amino acid). Correlation exists between insult level and phenotype function, affecting inflammation-related responses, including secondary pathogenesis.

It is of interest to compare the energies of the biochemical processes that result in the creation of additional NG2-glial cells and oligodendrocytes in healthy adults, as compared to the demyelination that occurs in response to injury with differentiation into myelinating oligodendrocytes, as an example.

6.  Concluding Summary

TBI is a devastating disease affecting millions of people worldwide, resulting in tens of thousands of deaths. Its link with brain energetics, and thus mitochondrial function, can help us better understand its progression and suggest clinical interventions that can ameliorate its most devastating symptoms.

## General Summary

We have presented and discussed current thinking about brain energetics, mitochondrial function, and traumatic brain injury. While each of these areas of study are in themselves vast with many unanswered questions, they are closely coupled and represent the complexities that researchers and clinicians need to be concerned with and aware of – that clinical interventions involve situations where serious ambiguities exist, and that most of our understanding of neurophysiological processes and mechanisms cross multiple scales, from subcellular to multicellular to full body scale, and are coupled across scales.

Nevertheless, specialized understanding at each scale needs to go forward independently of the other scales, and this can happen if the framework or control volume is defined properly. As progress is made at a particular scale, attempts can be made to couple the better-understood mechanisms to the other scales. Furthermore, an understanding in one discipline can lead to the same in a coupled discipline. For example, a better understanding of mitochondrial energetics will translate to more depth of understanding of TBI.

Here, brain energetics can be understood at the cellular level, how glia interact and support neurons, and how these cells convert and use energy from the nutrients ingested by the body. There are numerous mechanisms at the cellular level that need to be understood. At the subcellular level, numerous cell components work as a group to keep the cell functioning properly, where having sufficient energy is a core requirement, among others. At the center of that functioning are the mitochondria, whose internal workings are complex, as are how they interact with other organelles in the cell. Some of the internal workings of the cell are in response to inputs that originate from outside the cell. Similarly, these inner workings become inputs to mechanisms outside the cell. Some couplings and interactions operate over networks of cells, resulting in complex multicellular mechanisms that behave nonlinearly, switch on and off, have feedforward and feedback signals of biochemical, electrical and mechanical characteristics. Similarly, subcellular networks, in particular the mitochondrial and endoplasmic reticula, create couplings and exchange information and material.

All of these subcellular and cellular mechanisms are coupled to the larger scale of the entity they comprise – the human or animal body, or some smaller life form. That larger organism experiences and interacts with its environment, with its primary goal to survive and to ingest fuel in order to function, as a minimum. When the environment is damaging, as with TBI, the larger organism absorbs the insults at the large scale, converting and transmitting these all the way to the smallest scale of the organism, resulting in cellular and subcellular damage, triggering chemical responses coupled to mechanical behavior, resulting in dysfunction. This extremely complex nonlinear dynamic behavior is currently only understood in a piecemeal way.

These mechanisms at all scales "choose" among numerous progression paths, some that lead to dysfunction, in particular due to ineffective energy production. Understanding how these "choices" are made is based on being able to formulate models of the mechanisms that underlie all that has been described above. These "choices" and mechanisms are coupled across scales, even though it is possible to isolate parts of the processes for in-depth study. Is there some optimality in these "choices" or mechanisms? If so, then dysfunction and defects are sometimes optimal choices for the organism, and perhaps the optimizations are energy-dependent given the criticality of energy production and usage. There may not be any global optimizations, only local sub optimizations. Understanding these optimality decisions provides clues and eventually paths for clinical interventions and cures for some of humanity's most serious neurodegenerative diseases.

The review presented here is only a snapshot of current thinking and is, unavoidably, not a unique view of the subject matter. In the following section, a few final thoughts are presented.

## Highlights and Hypotheses

Dynamical equations, generally, can help researchers identify factors that may be difficult or impossible to determine experimentally. Dynamic models can be used to ascertain the impact of parameter uncertainties, as well as variations in clinical interventions. Dynamic models can help researchers examine cell behavior beyond those indicated by existing experiments, and optimally design additional experiments, perhaps reducing their number by relying on *in silico* trials.

There are numerous progression paths and possible initiators of neurodegenerative diseases where mitochondrial dysfunction can precede and possibly initiate, or occur concurrently, with an onset of pathologies. *It is hypothesized that the path "choices" made at the cellular and subcellular levels are (sub)optimal with respect to energetic functioning at scales within and beyond those of the organelle and the host cell; they can be (sub)optimal with respect to multiple cells, and have been evolutionarily conserved and refined.*

It is hypothesized that such optimal pathways can be identified via a variational framework (i.e., variational calculus) that *identifies possible pathways*, given morphological (mechanical), biochemical and metabolic constraints, and *"selects" the most likely pathway*. In some instances, clearly, the most likely pathway will be one that leads to dysfunction, for example, due to an imbalance between ROS production and their oxidation, or due to excessive fission. Constraints such as signaling mechanisms at all scales, cell and organelle morphology, feedback mechanisms, and imbalances of energetics and other intermediate products of mitochondrial functioning are a part of a possible formulation.

Variational theoretical models are able to represent dynamics and evolution, including the dissipative processes that exist in all biological systems. While homeostasis is a target, biological systems live dynamically around homeostasis. There are large energy flows about equilibrium by the constituent electrical, biochemical and mechanical processes. These dynamic flows are representable by governing differential equations derived via variational methods.

The bidirectional coupling between mitochondrial morphology (configurations) and bioenergetics suggests evolved optimality for efficiency, fast dynamics (fission and fusion), and high energy output in response to current needs.

Physiological and pathological processes result from biochemical, morphological and mechanical dynamics at more than one scale, and are likely suboptimal rather than optimal given the many competing molecular, organelle, and cellular interests. Mechanical aspects of cell and organelle dynamics cannot be decoupled form the biochemical mechanisms of physiological and pathological processes.

Astroglial metabolic networks linked by gap-junctions may represent an energetically efficient mechanism for glucose delivery to active neurons from a distal source rather than a nearby source that may already be in a high state of activity. Factors that can affect such mechanisms include capillary coverage by astrocytic endfeet, the density of glucose transporters in these endfeet, the astroglial metabolic machinery, and the strength of neuroglial interactions as witnessed by the density of receptors on astrocytes and the astrocytic coverage of neurons.

For TBI, a set of reduced order (simplified) theoretical models can be derived and correlated with available data, representing some of the key behaviors outlined above. Such equations can be used to extrapolate our understanding, and also be used to identify important targets for experimental investigations. Clearly identified assumptions can be removed as additional data becomes available and our understanding increases, resulting in a more complex and realistic theoretical model. Mechanical and topological (morphological) properties have been shown to influence biochemical and electrical processes. We hypothesize that the influence is significant, clearly during insult, but also subsequently as biochemical events unfold.

Given the structure of mitochondrial motility – the motor/adaptor complex, and the various possible anchor mechanisms – it may be possible, for example, to derive "equivalent mechanical" models that represent chemical bonds as mechanical forces, and perhaps suggest mechanical equivalents to chemical signaling, and thus an equivalent mechanical representation of trafficking.